\documentclass[conference,a4paper]{IEEEtran}
\usepackage{amssymb}
\usepackage{color}
\usepackage{multirow}
\usepackage{mathrsfs}
\usepackage{amsthm}

\usepackage{cite}
\ifCLASSINFOpdf
\usepackage[pdftex]{graphicx}
\graphicspath{{fig/}}
\DeclareGraphicsExtensions{.pdf,.jpeg,.png}
\else
\usepackage[dvips]{graphicx}
\graphicspath{{../eps/}}
\DeclareGraphicsExtensions{.eps}
\fi
\usepackage[cmex10]{amsmath}

\usepackage{array}

\usepackage[tight,footnotesize]{subfigure}
\usepackage{stfloats}
\usepackage{arydshln}
\usepackage{rotating}
\usepackage{microtype}
\def\gap{.5ex}
\abovedisplayskip\gap
\belowdisplayskip\gap
\abovedisplayshortskip\gap
\belowdisplayshortskip\gap

\begin{document}
%
\title{Optimization of Graph Based Codes for \\ Belief Propagation Decoding }

\author{\IEEEauthorblockN{Sachini Jayasooriya, Sarah J. Johnson, Lawrence Ong and Regina Berretta }
\IEEEauthorblockA{School of Electrical Engineering and Computer Science, 
University of Newcastle, Australia}}


%


\maketitle

\begin{abstract}
A low-density parity-check (LDPC) code is a linear block code described by a sparse parity-check matrix, which can be efficiently represented by a bipartite Tanner graph. The standard iterative decoding algorithm, known as belief propagation, passes messages along the edges of this Tanner graph. Density evolution is an efficient method to analyze the performance of the belief propagation decoding algorithm for a particular LDPC code ensemble, enabling the determination of a decoding threshold. The basic problem addressed in this work is how to optimize the Tanner graph so that the decoding threshold is as large as possible. We introduce a new code optimization technique which involves the search space range  which can be thought of as  minimizing randomness in differential evolution or limiting the search range in exhaustive search. This technique is applied to the design of good irregular LDPC codes and multi-edge type LDPC codes.
\end{abstract}


%

\section{Introduction}
The key objective of digital communications is to transmit information reliably from one point to another. With the introduction of iterative error correction codes (such as turbo, low-density parity-check and repeat-accumulate codes), error correction technology has become a vital means of achieving this aim in most current communication systems. A key performance measure of a coding scheme is its decoding threshold, which is the maximum noise level at which it can correct errors. In this paper, we design an efficient optimization technique to maximize the threshold of low-density parity-check (LDPC) codes.

In~\cite{Richardson2001-1,Gallager1963,Luby1999} a numerical technique, called Density Evolution (DE) was formulated to find the threshold of the belief propagation (BP) decoding algorithm for a given LDPC ensemble. An LDPC ensemble is the set of all LDPC codes with a particular property set, usually the degree distribution of their graphical (Tanner graph) representation. DE determines expected iterative decoding performance of a particular code ensemble by tracking the probability density function of Tanner graph edge messages through the iterative decoding process. This problem for the code designer is then to search for the ensemble with the best threshold from which a specific code may then be chosen. 

Multi-edge type LDPC (MET-LDPC) codes~\cite{Richardson2002multi} are a generalization of LDPC codes. Unlike standard LDPC ensembles which contain a single statical equivalence class of Tanner graph edges, in the multi-edge setting several edge classes can be defined and every node is characterized by the number of connections to edges of each class. The advantage of the MET generalization is greater flexibility in code structure and improved decoding performances.  

The code optimization of LDPC and MET-LDPC codes is a non-linear cost function maximization problem, where the DE threshold is the cost function and the Tanner graph structure and edge distribution gives the variables to be optimized.  In the majority of previous research in code optimization found in the literature, the optimization algorithm called Differential Evolution (Dif.E) has been applied to finding  good degree distributions for LDPC codes. This technique has been successfully applied to the design of good irregular LDPC codes for a range of channels~\cite{Richardson2001}. Shokrollahi and Sorn~\cite{Shokrollahi2005} used an improved version of Dif.E by proposing a new step called Discrete Recombination in order to increase the diversity of the new parameters in the search. Richardson and Urbanke~\cite{Richardson2002multi} suggested using hill-climbing method to optimize MET-LDPC codes. In our work, we develop a new code optimization technique to optimize codes more efficiently. This technique can be thought of as  minimizing the randomness in Dif.E~\cite{Richardson2001,Shokrollahi2005} or limiting the search space in ordinary exhaustive search and hill-climbing. This technique is then successfully applied to design good irregular LDPC codes and MET-LDPC codes.
 
In previous research of code optimization~\cite{Richardson2002multi,Richardson2001,Shokrollahi2005}, the structure of the LDPC and MET-LDPC Tanner graph is determined via trail and error or exhaustive search,  while only the edge distributions within a given structure are optimized. In this research, we propose a new nested method to optimize both the structure and edge distribution for LDPC and MET-LDPC codes. This is particularly important for MET-LDPC codes where, to date, it is not clear a priori which structures will be good.  

This paper is organized as follows. Section II briefly reviews the basic concepts of standard LDPC codes and MET-LDPC codes. In section III we review the code optimization problem for standard and MET-LDPC codes and discus our proposed code optimization technique. In Section IV we discuss the code optimization result obtained for several examples. Section V concludes the paper.

\section{Low-density Parity-check (LDPC) codes}

			\subsection{Standard LDPC codes}

As the name suggests, an LDPC code is a linear block code described by a sparse parity-check matrix. An LDPC parity-check matrix can be represented in graphical form by a Tanner graph. Suppose the LDPC parity-check matrix, $H$ has $N$ columns and $M$ rows; the corresponding Tanner graph consists of $N$ variable nodes, $M$ check nodes, and an edge for every non-zero entry in $H$. Each variable node represents a bit of the codeword while each check node represents a parity-check constraint of the code. Assuming $H$ is full rank, the code rate, r  is given by $(N-M)/N$. An LDPC code ensemble is typically specified by an edge degree distribution ($\lambda ,\rho$) from the perspective of Tanner graph edges:											\begin{equation}
					\label{eq : st_LDPC_lamda}
								\lambda(x) = \sum_{i=2}^{d_{(v,\text{max})}}\lambda_{i}x^{i-1} = \sum_{i\in \Lambda}\lambda_{i}x^{i-1}
					\end{equation}				
					\begin{equation}
					\label{eq : st_LDPC_roh}
								\rho(x) = \sum_{i=2}^{d_{(c,\text{max})}}\rho_{i}x^{i-1} = \sum_{i \in \Gamma}\rho_{i}x^{i-1}
					\end{equation}
where $\lambda_i$ (resp., $\rho_i$) is the fraction of edges that are connected to degree $i$ variable nodes (resp., check nodes) and $d_{(v,\text{max})}$ (resp., $d_{(c,\text{max})}$) is the maximum variable node degree (resp., check node degree). We let $\Lambda$ (resp., $\Gamma$)  be the set of $i$'s for non zero $\lambda_i$ (resp., $\rho_i$). 

\begin{figure}[!t]
							\centering
							\includegraphics[width=1.8in,natwidth=610,natheight=642]{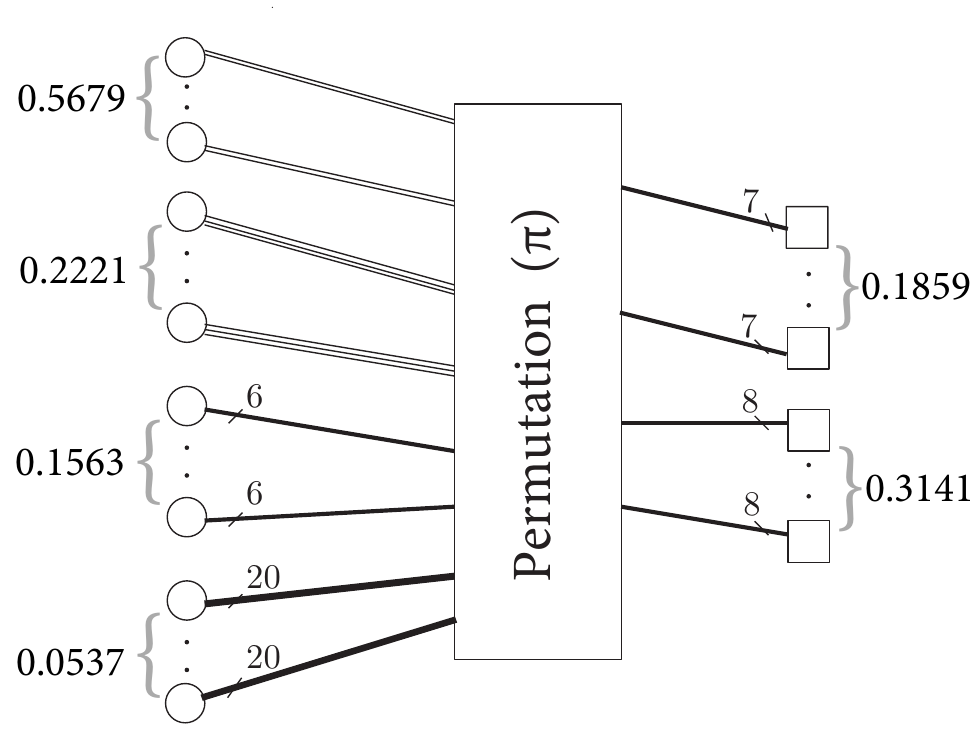}
							\caption{Bipartite Tanner graph representation of an example irregular LDPC code where \textquoteleft $\bigcirc$\textquoteright  (resp., \textquoteleft $\Box$\textquoteright) represents the variable nodes  (resp., check nodes). Number of nodes for different edge types are shown as fractions of the code length $N$ }
							\label{Fig.St_Tanner}
				  \end{figure}

The Tanner graph for a rate-half  irregular LDPC code is shown in Fig.~\ref{Fig.St_Tanner}, where $\Lambda$ = [2, 3, 6, 20], $\Gamma$ =[7, 8] and the degree distribution is given by $\lambda(x) = 0.2978x + 0.1747x^2 + 0.2459x^5 + 0.2816x^{19}$  and $\rho(x) = 0.3414x^6 + 0.6589x^7$.		
					
				\subsection{MET-LDPC codes}
					
MET-LDPC code ensembles are generally described based on a node-perspective, as opposed to the edge-perspective that is normally used for standard LDPC code ensembles. The MET-LDPC code ensemble can be specified through two multinationals associated to the variable and check nodes~\cite{richardson2008modern}.	
				
					\begin{equation}
					\label{eq : MET_LDPC_lamda}
								L(\boldsymbol{r},\boldsymbol{x}) = \sum L_{\boldsymbol{b},\boldsymbol{d}} \boldsymbol{r}^{\boldsymbol{b}} \boldsymbol{x}^{\boldsymbol{d}}
					\end{equation}					
					\begin{equation}
					\label{eq : MET_LDPC_roh}
								R(\boldsymbol{x}) = \sum R_{\boldsymbol{d}} \boldsymbol{x}^{\boldsymbol{d}}
					\end{equation}
where $\boldsymbol{b}, \boldsymbol{d}, \boldsymbol{r}$ and $\boldsymbol{x}$ are vectors defined as follows. Let $m_e$ denote the number of edge types used in the graph ensemble and  $m_r$ denote the number of different channels over which a bit may be transmitted. Let the vector $\boldsymbol{d} = [d_1, \ldots, d_{m_e}]$ be a multi-edge degree and $\boldsymbol{b} = [b_0, \ldots,b_{m_r}]$ be a received degree where $b_0$ is associated with punctured variables (variables not transmitted to the receiver). The vector of variables corresponding to the edge distributions is denoted by $\boldsymbol{x} = [x_1, \ldots, x_{m_e}]$ and $\boldsymbol{x}^{\boldsymbol{d}} = \prod_{i=1}^{m_e} x_i^{d_i}$. The vector of variables corresponding to the received distributions is denoted by $\boldsymbol{r} = [r_0, \ldots, r_{m_r}]$ and $\boldsymbol{r}^{\boldsymbol{b}} =\prod_{i=0}^{m_r} r_i^{b_i}$. $L_{\boldsymbol{b},\boldsymbol{d}}, R_{\boldsymbol{d}}$ are non-negative reals corresponding to the fraction of variable nodes with type ($\boldsymbol{b}, \boldsymbol{d}$) and the fraction of check nodes with type ($\boldsymbol{d}$) in the graph respectively. In this research, all the received variables are transmitted through a single link (i.e $m_r$ = 1). Hence for un-punctured variables in the codeword (i.e $b_0$ = 0, $b_1$ = 1) $\boldsymbol{b}$ = [0, 1] and for punctured variables(i.e $b_0$ = 1, $b_1$ = 0) $\boldsymbol{b}$ = [1, 0].
\section{Problem Statement}
We can determine the decoding threshold for a given LDPC code ensemble defined by its degree distribution pair ($\lambda,\rho$) via DE~\cite{richardson2008modern}. Our task is to find the degree distribution pair which yields the largest possible threshold. This a is non-linear cost function maximization problem.
			\subsection{Problem statement for standard LDPC codes}
					On the binary erasure channel (BEC) the optimization problem is as follows. On other channels an appropriate DE function, or a suitable approximation, is used for (5).\\
													
					For a fixed  code rate, r and maximum number of decoder iterations, $l$, where r $\in [0, 1]$ and $l\in \{1, 2, \ldots\}$, 
							
					Max $\epsilon^{*},$     \hspace{0.5cm} $0 \leq \epsilon^{*} \leq 1$
					
					Subject to:  \hspace{0.25cm} $f(\epsilon^{*},  l,  \lambda,  \rho) = 0$
					
					where $f(\epsilon^{*}, l, \lambda, \rho)$ is a recursive function described by																											
					\begin{align}
									f(\epsilon^{*}, l, \lambda, \rho) &= \epsilon^{(l)} = \epsilon^{*} \lambda (1 - \rho(1 - \epsilon^{(l-1)})) \\
									\epsilon^{(0)} &= \epsilon^{*}
					\end{align}	
			
					Variables: $\lambda,  \rho$\\										
					where $\lambda(x)$ and $\rho(x)$ are given by (\ref{eq : st_LDPC_lamda}) and (\ref{eq : st_LDPC_roh}) respectively.
												
					Constraints:
									\begin{align} 
														\sum_{i=2}^{d_{(v,\text{max})}}\lambda_{i} &= 1  \\
														\sum_{i=2}^{d_{(c,\text{max})}} \rho_{i} &= 1\\
														\frac{\sum_{i=2}^{d_{(c,\text{max})}} \rho_{i}/i}{\sum_{i=2}^{d_{(v,\text{max})}} \lambda_{i}/i} &= 1 - \text{r}\\
														\lambda^{'}(0)\rho^{'}(1) &< \frac{1}{\epsilon^{*}}													
									\end{align}
					where $\lambda^{'}(0) = \frac{d}{dx}\lambda(x)\big|_{x=0}$ and $\rho^{'}(1)= \frac{d}{dx}\rho(x)\big|_{x=1}$
							
					\vspace{0.1cm}
				
					Traditionally, the optimization problem is considered for a fixed $\Lambda$ and $\Gamma$ chosen via trail and error or intuition. That is the allowed degrees (for which $\lambda_i$ and $\rho_i$ are non-zero) are fixed in advance. Here we include the allowed degrees as variables in the optimization. This allows later generalization to MET-LDPC codes where the choice of $\Lambda$ and $\Gamma$ is not straight forward.	
								
					In this case we add constraints to restrict the number of non-zero entries in $\lambda(x)$ and $\rho(x)$ and their maximum range as follows:
									\begin{align}
														|\Lambda| &\leq \Lambda_{\text{max}}\\
														|\Gamma| &\leq \Gamma_{\text{max}}\\
														\text{Max}(\Lambda) &\leq d_{(v,\text{max})}\\
														\text{Max}(\Gamma) &\leq d_{(c,\text{max})}
									\end{align}			
									
				\subsection{Problem statement for MET-LDPC codes}			
								Here we show the problem statement for a MET-LDPC code. In this case $\boldsymbol{\epsilon}^{(l)}$ is a vector $\boldsymbol{\epsilon}^{(l)} = [\epsilon_{1}^{(l)}, \ldots , \epsilon_{m_e}^{(l)}]$  where $m_e$ is the number of edge classes and $\boldsymbol{\epsilon}^{(0)}_i = \epsilon^{*}$ for all $i$.\\
			
					For a fixed  code rate, r and maximum number of decoder iterations, $l$, where r $\in [0, 1]$ and $l\in \{1, 2, \ldots\}$,  	
										
					Max $\epsilon^{*}$,  \hspace{0.5cm}  $0 \leq \epsilon^{*} \leq 1$
					
					Subject to:  \hspace{0.25cm} $f(\boldsymbol{\epsilon}^{(0)}, l, \boldsymbol{\lambda}, \boldsymbol{\rho}) = 0$
					
					where $f(\boldsymbol{\epsilon}^{(0)}, l, \boldsymbol{\lambda}, \boldsymbol{\rho})$ is a recursive function described by																						
				  \begin{align} 
							f(\boldsymbol{\epsilon}^{(0)}, l, \boldsymbol{\lambda,} \boldsymbol{\rho}) &= \boldsymbol{\epsilon}^{(l)} = \boldsymbol{\lambda} (\boldsymbol{\epsilon}^{(0)}, \boldsymbol{\rho}(\boldsymbol{\epsilon}^{(l-1)}))							
					\end{align}
																		
					where
					\begin{equation}
										\boldsymbol{\lambda}(\boldsymbol{r},\boldsymbol{x}) = \left(\frac{L_{x_1}(\boldsymbol{r},\boldsymbol{x})}{L_{x_1}(\boldsymbol{1},\boldsymbol{1})}, \ldots,  \frac{L_{x_{m_e}}(\boldsymbol{r},\boldsymbol{x})}{L_{x_{m_e}}(\boldsymbol{1},\boldsymbol{1})} \right)
					\end{equation}					
					\begin{equation}
										\boldsymbol{\rho}(\boldsymbol{x}) = \left(\frac{R_{x_1}(\boldsymbol{x})}{R_{x_1}(\boldsymbol{1})}, \ldots,  \frac{R_{x_{m_e}}(\boldsymbol{x})}{R_{x_{m_e}}(\boldsymbol{1})} \right)
					\end{equation}
														
					Constraints:
									\begin{align} 
														L_{x_i}(\boldsymbol{r},\boldsymbol{x}) &= \frac{d}{d_{x_i}} L(\boldsymbol{r},\boldsymbol{x})\\
														R_{x_i}(\boldsymbol{x}) &= \frac{d}{d_{x_i}} R(\boldsymbol{x})\\
														L_{b_1}(\boldsymbol{1},\boldsymbol{1}) &= 1 \\
														L(\boldsymbol{1},\boldsymbol{1}) - R(\boldsymbol{1}) &= \text{r}\\
														L_{x_i}(\boldsymbol{1},\boldsymbol{1}) &= R_{x_i}(\boldsymbol{1}) 									
									\end{align}							
	       where $L(\boldsymbol{r},\boldsymbol{x})$ and $R(\boldsymbol{x})$ are given by (\ref{eq : MET_LDPC_lamda}) and (\ref{eq : MET_LDPC_roh}) respectively and $\boldsymbol{1}$ denotes a vector of all 1's with the length determined by the context. \\
				
				We will consider these optimization problems using a range of existing non-convex optimization methods and a proposed method.

			\subsection{Proposed Code optimization Algorithm: Adaptive Range Method}
			
			The Adaptive Range (AR) method forms the next set of points for evaluation as a random selection of the set of points close to the current optimum over the search space. The size of the search space is adapted as the algorithm progresses. The algorithm is as follows:

Inputs are:
\begin{itemize}
	\item{Optimization parameters: Population size (NP),	range multiplier (RM), The tolerance on the best vector ($\delta$)} 
	\item{Tanner graph limits: $\Lambda_{\text{max}},  \Gamma_{\text{max}},  d_{(v,\text{max})},  d_{(c,\text{max})}$}
	\item{Decoder limits: Maximum number of iterations ($l$)}
\end{itemize}

\vspace{0.25cm}

\begin{enumerate}
			\item {\textbf{Initialization} $-$
						For the first generation ($G=0$), choose NP length-$E$ vectors $\boldsymbol{P}_{i,G}$, $i = 0,1, \dotsc ,\text{NP}-1$, using the Queen’s move strategy~\cite{Shokrollahi2005} where $E$ is the number of free elements of the degree distribution pair ($\lambda, \rho$).}
			\item{\textbf{Threshold} $-$
						For each vector $\boldsymbol{P}_{i,G}$, run DE  for the given number of iterations and record the ensemble threshold ($\epsilon^{*}$). Then 
						select the vector with largest threshold ($\boldsymbol{P}_{\text{Best},G}$, $\epsilon_{\text{Best},G}$) and the vector with next largest threshold ($\boldsymbol{P}_{\text{NextBest},G}$, $\epsilon_{\text{NextBest},G}$).}
			\item{\textbf{Random local search} $-$
						For the next generation  $G+1$, new vectors are generated according to the following scheme. For each $\boldsymbol{P}_{i,G+1}$, $i = 0,1, \dotsc ,\text{NP}-1$, randomly choose each of its element that is at most SR away from that of the current best vector. Note that each element must lie between zero and one.							
              \begin{multline}
								\boldsymbol{P}_{i,G+1}^{j} = \text{rand}[\text{max}(\boldsymbol{P}^{j}_{\text{Best},G} - \text{SR}, 0),  \\ 
                                                               \text{min}(\boldsymbol{P}^{j}_{\text{Best},G} + \text{SR}, 1)]
							\end{multline}
							\begin{equation}
								\boldsymbol{P}_{0,G+1}^{j} = \boldsymbol{P}_{\text{Best},G}^{j}
							\end{equation}
							\begin{equation}
								1 \leq j \leq E
							\end{equation}}
					\vspace{-1em}						
			\item{\textbf{Recalculation of search range} $-$
						Recalculate  SR when $\epsilon_{\text{Best},G} - \epsilon_{\text{Best},G-1} < \delta$ 
						\begin{align}
								\text{SR} &= \text{RM} \times \text{max}_{j \in 1..E} ( |\boldsymbol{P}_{\text{Best},G}^{j} - \boldsymbol{P}_{\text{NextBest},G}^{j} |,0.0001)
						\end{align}}
				\vspace{-1em}				
			\item{\textbf{Stopping criterion} $-$
						Halt if there is no improvement in threshold after three iterations.  Otherwise return to Step 2.}		
\end{enumerate}

\section{Optimization of standard LDPC codes}

		\subsection{Optimization of ($\lambda, \rho$) for a given set of allowed degrees}
		
				 \begin{figure}[!b]
							\centering
							\includegraphics[width=2.5in,natwidth=610,natheight=642]{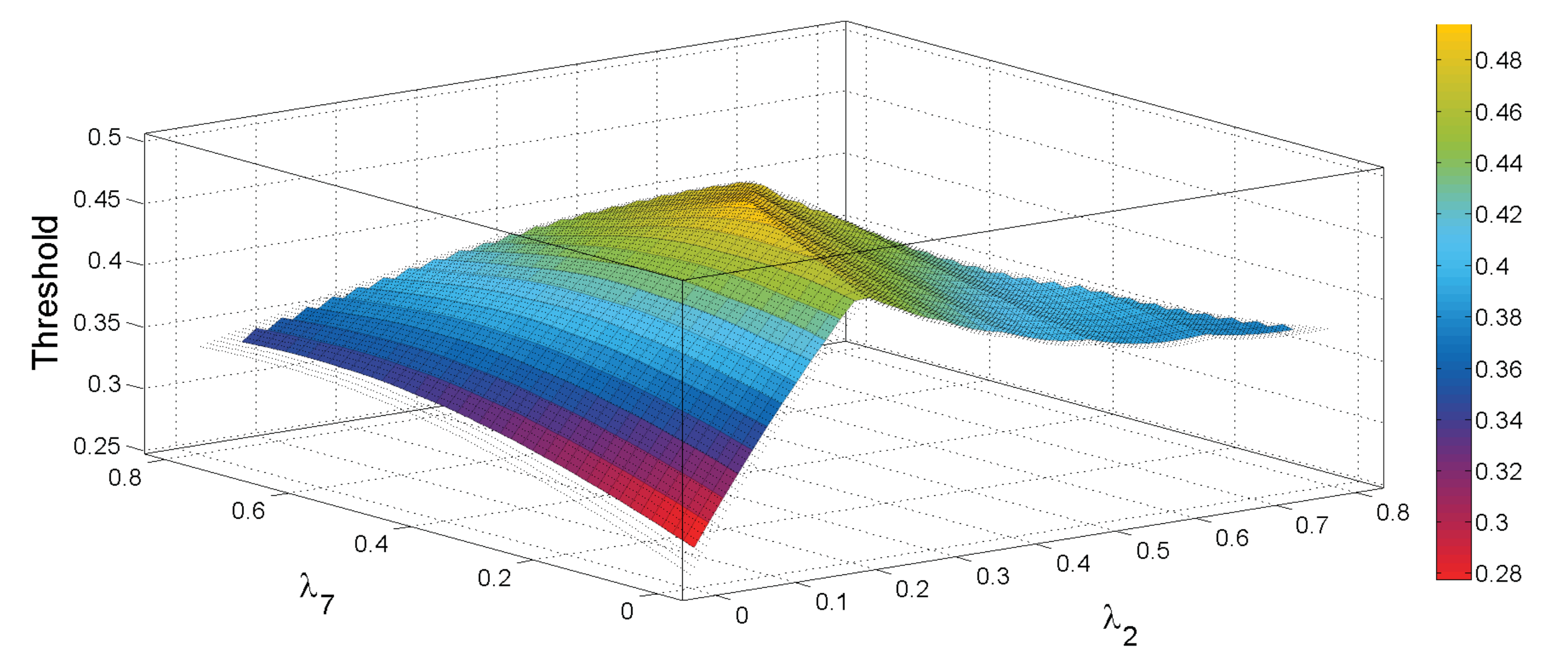}
							\caption{DE cost function for fixed $\Lambda$ = [2, 3, 7, 25], where $\lambda_3$ is fixed to 0.2024 and $\lambda_{20}$ given by (7)}
							\label{Fig.cost function}
				  \end{figure}

					\begin{table*}[!t]
							\renewcommand{\arraystretch}{1.3}
							\caption{The degree distributions found using 3 different optimization techniques for rate-half codes considered in~\cite{Shokrollahi2005}}
							\label{Table : Some good degree distributions}
							\centering
							\begin{tabular}{l |c| c| c |c |c |c| c| c| c||c| c| c |  }
							\cline{2-13}
							{}  &  \multicolumn{3}{c|}{$\boldsymbol{\Lambda = [2, 3, 6, 20]}$} & \multicolumn{3}{c|}{$\boldsymbol{\Lambda = [2, 3, 7, 25]}$} & \multicolumn{3}{c||}{$\boldsymbol{\Lambda = [2, 3, 7, 30]}$} & \multicolumn{3}{c|}{$\boldsymbol{|\Lambda| \leq 4, \text{Max}(\Lambda)\leq 30 }$}  \\
							\cline{2-13}
							{} & {\textbf{AR}} & {\textbf{Dif.E}} & {\textbf{Dif.E.R}} &  {\textbf{AR}} & {\textbf{Dif.E}} & {\textbf{Dif.E.R}} &  {\textbf{AR}} & {\textbf{Dif.E}} & {     
							\textbf{Dif.E.R}} & {\textbf{AR}} & {\textbf{Dif.E}} & {\textbf{RS}}\\
							\cline{2-13}
							\cline{1-13}
										\multicolumn{1}{|l|}{$\lambda_2$}	&	0.2962	&	0.2985	&	0.2987	&	0.2774	&	0.2750	&	0.2770	&	0.2621	&	0.2630	&	0.2636	&	0.2610	&	0.2672	&	0.2860	\\
										\multicolumn{1}{|l|}{$\lambda_3$}	&	0.1749	&	0.1740	&	0.1741	&	0.2020	&	0.2040	&	0.2025	&	0.1816	&	0.1810	&	0.1801	&	0.1832	&	0.1758	&	-	\\
										\multicolumn{1}{|l|}{$\lambda_4$}	&	-	&	-	&	-	&	-	&	-	&	-	&	-	&	-	&	-	&	-	&	-	&	0.3326	\\
										\multicolumn{1}{|l|}{$\lambda_6$}	&	0.2418	&	0.2485	&	0.2489	&	-	&	-	&	-	&	-	&	-	&	-	&	-	&	-	&	-	\\
										\multicolumn{1}{|l|}{$\lambda_7$}	&	-	&	-	&	-	&	0.2626	&	0.2560	&	0.2610	&	0.2670	&	0.2690	&	0.2706	&	0.2640	&	0.2797	&	-	\\
										\multicolumn{1}{|l|}{$\lambda_{13}$}	&	-	&	-	&	-	&	-	&	-	&	-	&	-	&	-	&	-	&	-	&	-	&	0.1834	\\
										\multicolumn{1}{|l|}{$\lambda_{20}$}	&	0.2872	&	0.2790	&	0.2784	&	-	&	-	&	-	&	-	&	-	&	-	&	-	&	-	&	-	\\
										\multicolumn{1}{|l|}{$\lambda_{25}$}	&	-	&	-	&	-	&	0.2580	&	0.2650	&	0.2595	&	-	&	-	&	-	&	-	&	-	&	-	\\
										\multicolumn{1}{|l|}{$\lambda_{30}$}	&	-	&	-	&	-	&	-	&	-	&	-	&	0.2893	&	0.2870	&	0.2856	&	0.2918	&	0.2772	&	0.1980	\\
										\multicolumn{1}{|l|}{$\rho_{7}$}	&	0.3094	&	0.3533	&	0.3571	&	0.1083	&	0.0748	&	0.1026	&	-	&	-	&	-	&	-	&	-	&	-	\\
										\multicolumn{1}{|l|}{$\rho_{8}$}	&	0.6976	&	0.6467	&	0.6429	&	0.8917	&	0.9252	&	0.8974	&	0.6171	&	0.6338	&	0.6418	&	0.6036	&	0.6912	&	0.8872	\\
										\multicolumn{1}{|l|}{$\rho_{9}$}	&	-	&	-	&	-	&	-	&	-	&	-	&	0.3829	&	0.3662	&	0.3582	&	0.3964	&	0.3088	&	0.1128	\\
										\hline	
										\multicolumn{1}{|l|}{\textbf{$\epsilon^{*}$}}	&	\textbf{0.4939}	&	\textbf{0.4940}	&	\textbf{0.4940}	&	\textbf{0.4949}	&	\textbf{0.4949}	&	\textbf{0.4949}	&	\textbf{0.4955}	&	\textbf{0.4955}	&	\textbf{0.4955}	&	\textbf{0.4955}	&	\textbf{0.4954}	&	\textbf{0.4915}	\\
										\multicolumn{1}{|l|}{(AVG)}	&	0.4935	&	0.4939	&	0.4938	&	0.4948	&	0.4949	&	0.4948	&	0.4954	&	0.4955	&	0.4955	&	0.4955	&	0.4954	&	0.4915	\\
										\multicolumn{1}{|l|}{(SD)}	&	2.2E-4	&	9.8E-5	&	6.0E-5	&	9.2E-5	&	0	&	1.6E-4	&	1.8E-4	&	5.5E-17	&	5.5E-17	&	5.5E-17	&	0	&	5.5E-17	\\	
										\hline	
										\multicolumn{1}{|l|}{NP}	&	50	&	50	&	50	&	100	&	100	&	100	&	100	&	100	&	100	&	50	&	50	&	50	\\
										\multicolumn{1}{|l|}{NOG}	&	18	&	25	&	14	&	14	&	21	&	14	&	15	&	21	&	14	&	5860	&	3430	&	7036	\\
										\multicolumn{1}{|l|}{NTT}	&	900	&	1250	&	700	&	1400	&	2100	&	1400	&	1500	&	2100	&	1400	&	286778	&	171046	&	348626	\\
										\multicolumn{1}{|l|}{NFE}	&	51	&	42	&	700	&	42	&	34	&	1400	&	37	&	29	&	1400	&	22345	&	11702	&	21729	\\
										\hline	
										\multicolumn{1}{|l|}{CPU.T}	&	\textbf{71.277}	&	\textbf{87.002}	&	\textbf{2062.1}	&	\textbf{63.586}	&	\textbf{82.072}	&	\textbf{3349.0}	&	\textbf{63.960}	&	\textbf{84.100}	&	\textbf{3787.4}	&	\textbf{12744}	&	\textbf{6846.9}	&	\textbf{13027}	\\
										\multicolumn{1}{|l|}{(AVG)}	&	74.785	&	155.50	&	2012.9	&	86.592	&	118.54	&	3669.8	&	70.664	&	102.53	&	3853.8	&	13483	&	7573.1	&	15282	\\
										\multicolumn{1}{|l|}{(SD)}	&	14.382	&	40.313	&	136.92	&	16.662	&	28.435	&	180.32	&	13.278	&	16.804	&	61.308	&	801.12	&	688.07	&	1184.9		\\		
												
							\hline							    
										\multicolumn{13}{|l|}{$\epsilon^{*} -$Ensemble Threshold, NOG $-$ No of Generations before algorithm halted, NTT $-$ Number of times we check the degree distribution at }\\
										\multicolumn{13}{|l|}{ best threshold so far (Number  of candidates considered), NFE $-$ Number of times we find a new Threshold (New degree distribution   }\\
										\multicolumn{13}{|l|}{improves all previous best), CPU.T $-$ CPU time in seconds, AVG $-$ Mean value, SD $-$ Standard deviation} \\
										
							\hline
						\end{tabular}
					\end{table*}
	\vspace{-0.08cm}	
	
					 To find the best optimization algorithm of ($\lambda, \rho$) given ($\Lambda, \Gamma$) fixed, we compare our proposed algorithm, AR method with Dif.E~\cite{Richardson2001} and Differential Evolution with Recombination (Dif.E.R)~\cite{Shokrollahi2005}.  We considered  three optimization problems from~\cite{Shokrollahi2005}. In all cases, initialization was via Queen’s move strategy~\cite{Shokrollahi2005} and stopping criterion was set as halt if there is no improvement in threshold after three generations.  The population size at each generation was fixed at 50 (case 1) or 100 (case 2 and 3). For all three optimization problems, we set RM=0.5 in the AR method.  Table~\ref{Table : Some good degree distributions} shows our results and the table entries give the degree distribution and simulation parameters for the best case over 10 trials. The numbers in brackets give the average and standard deviation for $\epsilon^{*}$ and simulation time (CPU.T) over the 10 trials. While all three algorithms returned the same optimal threshold up to three significant decimal points, the AR method was significantly quicker. The superiority of the AR method, in this example, is due to the shape of the DE cost function for fixed ($\Lambda, \Gamma$)  as shown in Fig.~\ref{Fig.cost function}. Although rounding off the threshold to a fixed number of significant figures does result in multiple points with the same threshold, there are otherwise no local maxima to trap the optimization algorithm. The AR method, which is a type of local search optimization technique works well in the absence of multiple local maxima.

			\subsection{Generalization of degree distribution}
						
				Next, we jointly optimized ($\Lambda, \Gamma$) and ($\lambda,\rho$) using the AR method, Dif.E and Random Search (RS) for the ($\Lambda, \Gamma$) variables with an inner optimization of AR for the ($\lambda_i,\rho_i$) variables. When including the allowed degrees in the optimization, the task  now  is to solve an integer optimization, as the node degrees can only be integer values. In the AR method, optimization of ($\lambda,\rho$) was done using RM and SR (initial setting) as described in section IV.A. For optimization of ($\Lambda, \Gamma$), 1 and 15 was selected for RM and SR (initial setting) respectively.  To  compare the three optimization algorithms, we consider LDPC codes with rate-half, $|\Lambda| = 4$, $|\Gamma| = 2$  and constraint on the maximum bit node degree of $d_{v,\text{max}} =$ 30. In all cases $\lambda_i,\rho_i$ were initialized with the Queens move method. The last three columns of Table~\ref{Table : Some good degree distributions} show our results. The combined optimization using either AR or Dif.E returns the same ($\Lambda, \Gamma$) set as found in~\cite{Shokrollahi2005}. Thus blind optimization returns the same degree set found in~\cite{Shokrollahi2005} via intuition and trail and error. The AR method was significantly faster than the Dif.E method and the RS method failed to achieve the best degree combination in all 10 trails. AR and Dif.E algorithms return the same threshold value with an accuracy of three significant decimal points for the same ($\Lambda, \Gamma$), although the ($\lambda, \rho$) sets are sightly different. This is because there can be more than one  ($\lambda, \rho$) set for each ($\Lambda, \Gamma$) that obtains the same threshold when it is evaluated within three  significant figures.

\section{Optimization of MET-LDPC codes}
In this section we apply our optimization techniques to MET-LDPC ensembles where the combined optimization technique is particularly useful to explore the wide range of possible structures.

\begin{figure}[!t]
										\centering
													\includegraphics[width=1.8in,natwidth=610,natheight=642]{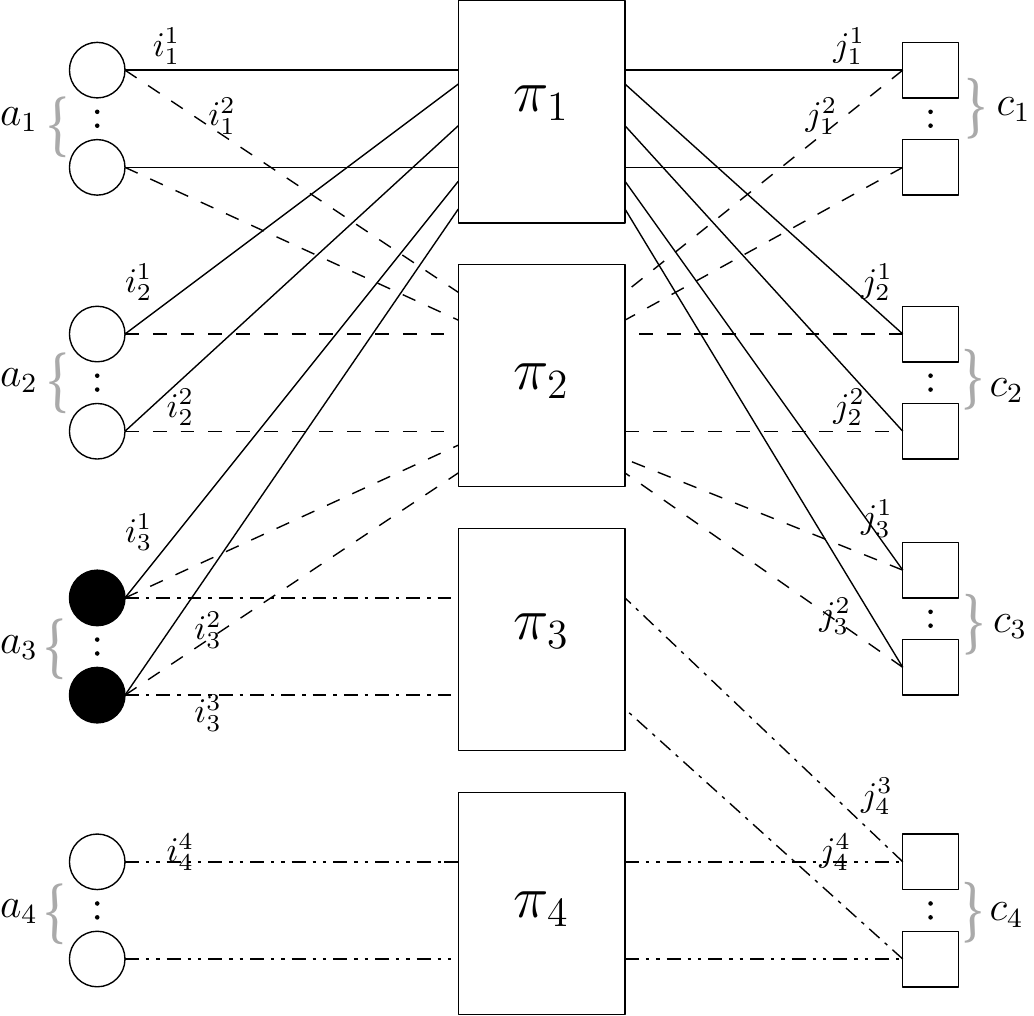}
													\vspace{-1em}
										\caption{Bipartite Tanner graph representation of 4-edge class MET-LDPC code where subscript (resp., superscript) letters denote the node class (resp., edge class).\textquoteleft $\bullet$\textquoteright represent punctured variable nodes.}
										\label{Fig.MET_LDPC}
								\end{figure}

\begin{figure}[!t]
											\centering
											\includegraphics[width=3in,natwidth=610,natheight=642]{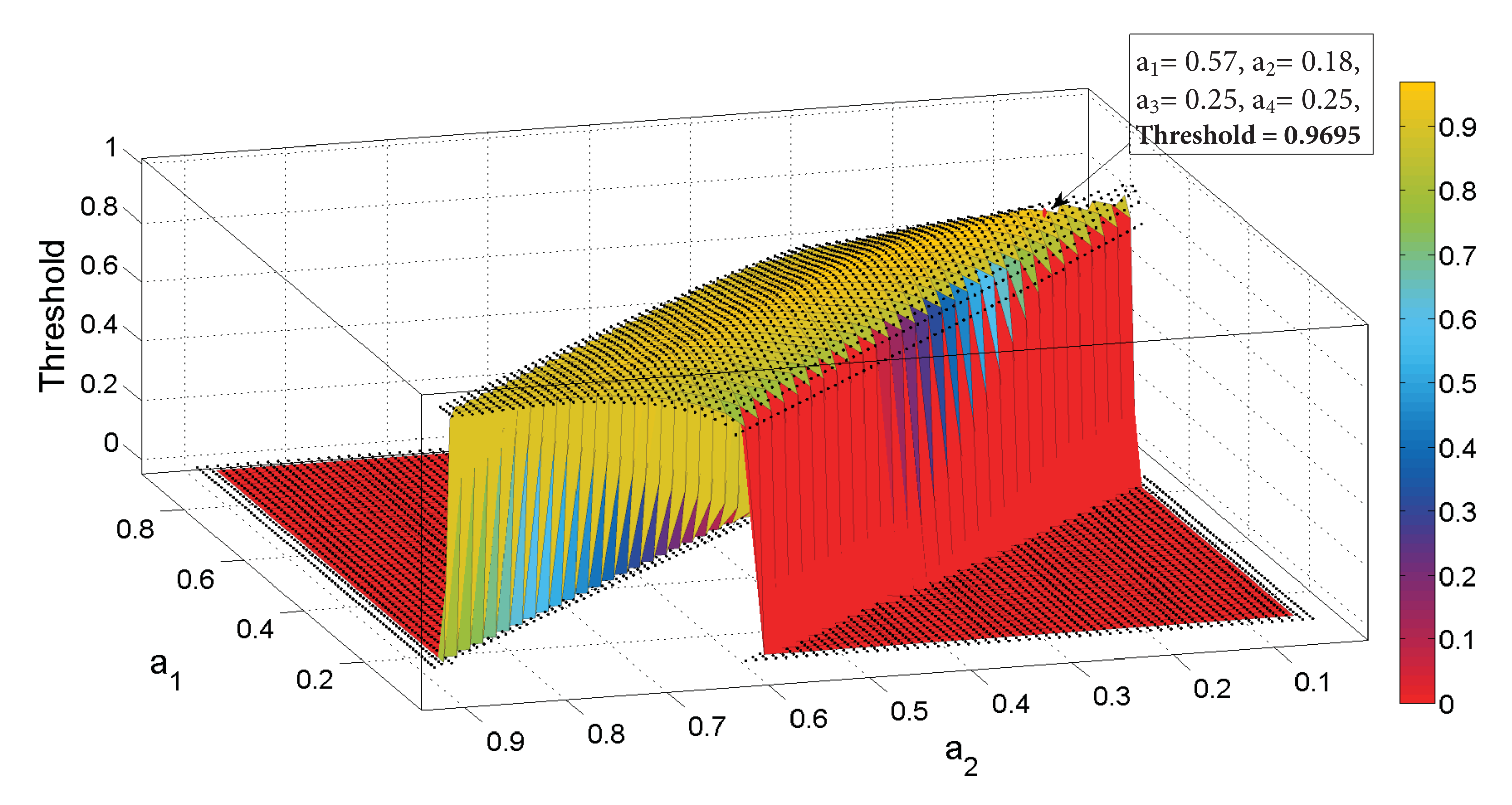}
											\vspace{-1em}
											\caption{MET-DE cost function for optimization of degree distribution for fixed $L(\boldsymbol{r},\boldsymbol{x}) = a_1r_1x_1^2 + a_2r_1x_1^3 + a_3r_0x_2^3x_3^3 + a_4r_1x_4 $ (i.e $\Lambda_1 = [2, 3, 0, 0], \Lambda_2 = [0, 0, 3, 0], \Lambda_3 = [0, 0, 3, 0], \Lambda_4 = [0, 0, 0, 1]  $) where $a_3= a_4$ and $a_3$ is given by (20)}
											\label{Fig.MET inner cost function}
								\end{figure}

Similarly to standard LDPC codes we found that, for the inner optimization (i.e., optimizing the node fractions, denoted as $a_i$ and $c_i$ in Fig.~\ref{Fig.MET_LDPC}, for a fixed code structure, denoted as $i_k^\ell$ and $j_k^\ell$ in the same figure), the AR method returns a higher optimal threshold than Dif.E. The MET-DE cost function when allowed degrees are fixed, has a global maximum and no other local maxima. An example of this cost function for the 4-edge class MET-LDPC code given in Table VI of~\cite{Richardson2002multi} on a BI-AWGN channel is shown in Fig.~\ref{Fig.MET inner cost function}.

\subsection{Generalization of degree distribution}

\begin{figure}[!t]
											\centering
											\includegraphics[width=3in,natwidth=610,natheight=642]{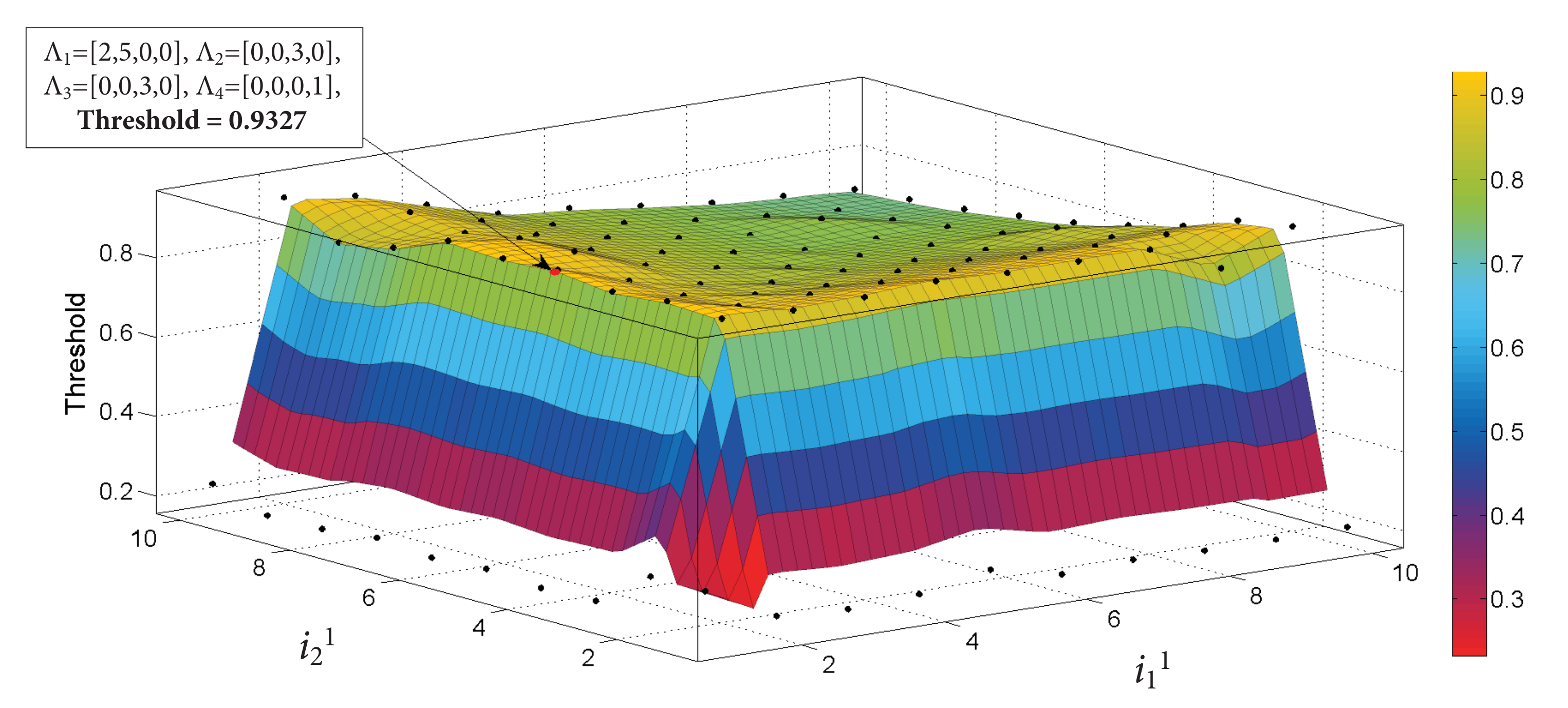}
											\vspace{-1em}
											\caption{MET-DE cost function for optimization of MET code structure where $\Lambda_1 = [i^1_1, i^1_2, 0, 0] , \Lambda_2 = [0, 0, 3, 0], \Lambda_3 = [0, 0, 3, 0], \Lambda_4 = [0, 0, 0, 1]  $  }
											\label{Fig.MET outer cost function}					
				        \end{figure}

To examine the properties of the MET-DE cost function for the generalization of the MET-LDPC code structure, we consider MET-LDPC codes with four variable node classes, maximum allowable variable node degree of 10 and edge class 4 containing degree-1 variable nodes as in the example in~\cite{Richardson2002multi}. The MET-DE  cost function for this situation has local maxima. Fig.~\ref{Fig.MET outer cost function} shows an example of this cost function as the node degree of the class 1 and class 2 variable nodes are varied. In this scenario Dif.E outperforms the AR method by escaping local maxima which is reflected in the performance of each algorithm as shown in Table~\ref{Table : MET optimization} . Hence we propose that,  MET-LDPC codes be optimized using the AR method as the MET degree distribution optimization technique and Dif.E  to optimize the MET code structure.  Through the joint optimization we gain a threshold improvement of 0.007 compared to the reference MET-LDPC code (Table VI of~\cite{Richardson2002multi}) which has a threshold of $\sigma^{*}$=0.9682 over the BI-AWGN channel.  Here we see the benefit of MET-LDPC codes as standard LPDC codes require a maximum node degree of 85 to obtain a threshold this high~\cite{chung2000thesis}.
			\vspace{-0.1cm}
		
									\begin{table}[!t]
										\renewcommand{\arraystretch}{1.3}
										\caption{Optimization of 4-edge class rate-half MET-LDPC codes on the BI-AWGN channel}\label{Table : MET optimization}
										\centering
										\begin{tabular}{|c | c| c |c |c |c |c | c| c|}
											\hline
											\multicolumn{4}{|c|}{$\boldsymbol{d}$ } & \multicolumn{2}{c|}{$\boldsymbol{b}$ } &\multicolumn{3}{c|}{$\boldsymbol{L_{b,d}}$}\\
												\hline															
												{$\boldsymbol{i^1}$}	&	{$\boldsymbol{i^2}$} &	{$\boldsymbol{i^3}$} &	{$\boldsymbol{i^4}$} &	{$\boldsymbol{b_0}$} &	{$\boldsymbol{b_1}$} & {\textbf{Ref.}}&	{\textbf{AR}}	&	{\textbf{Dif.E}} \\
												\hline
												2 & 0 & 0 & 0 & 0 & 1 & 0.5 & 0.5658  & 0.3058 \\
												3 & 0 & 0 & 0 & 0 & 1 & 0.3 & 0.1878  & -\\
												2 & 1 & 1 & 0 & 0 & 1 &   - &	 -      & 0.0910\\
												0 & 3 & 3 & 0 & 1 & 0 & 0.2 & 0.2464  &  -\\
												2 & 3 & 3 & 0 & 1 & 0 & -		&	 -      & 0.4021\\
												0 & 0 & 0 & 1 & 0 & 1	& 0.2 & 0.2464  & 0.6031\\														
												\hline															
												\multicolumn{4}{|c|}{$\boldsymbol{d}$}	& \multicolumn{2}{c|}{}	&	\multicolumn{3}{c|}{$\boldsymbol{R_{d}}$}	\\	
												\cline{1-4}															
												{$\boldsymbol{j^1}$}	&	{$\boldsymbol{j^2}$}	&	{$\boldsymbol{j^3}$} &	{$\boldsymbol{j^4}$} &	\multicolumn{2}{c|}{} &  \multicolumn{3}{c|}{}	\\	
												\hline												
												3 & 1 & 0 & 0 & \multicolumn{2}{c|}{}   &-   & 0.2609 & - \\
												3 & 2 & 0 & 0 & \multicolumn{2}{c|}{}   &-   & 0.0441 & -      \\
												5 & 4 & 0 & 0 & \multicolumn{2}{c|}{}   &-   & -      & 0.1958 \\
												6 & 4 & 0 & 0 & \multicolumn{2}{c|}{}   &-   & -      & 0.0017  \\
												6 & 5 & 0 & 0 & \multicolumn{2}{c|}{}   &-   & -      & 0.1015  \\
												2 & 2 & 1 & 0 & \multicolumn{2}{c|}{}   &0.4 & -      & -     \\
												2 & 1 & 2 & 0 & \multicolumn{2}{c|}{}   &0.1 & -      &  -     \\
												0 & 0 & 3 & 1 & \multicolumn{2}{c|}{}   &0.2 & 0.2464 & -  \\
												0 & 0 & 2 & 1 & \multicolumn{2}{c|}{}   &-   & -      & 0.6031  \\									
												\hline																														
												\multicolumn{6}{|c|}{$\sigma^{*}$}	&	\textbf{0.9682}	&		\textbf{0.9692}  &	\textbf{0.9754}		\\														
												\multicolumn{6}{|c|}{(AVG)}	        &			-           &   0.9544  & 	0.9734	\\															
												\multicolumn{6}{|c|}{(SD)}	        &				-         &   0.0146  & 0.0024	\\		
												\hline															
												\multicolumn{6}{|c|}{CPU.T $\times 10^3$}&			-	   &	\textbf{6.0961}	  &	\textbf{110.28}			\\															
												\multicolumn{6}{|c|}{(AVG $\times 10^3$)}	           &			-	          &	7.9262	  &	 211.24 	\\																			
												\multicolumn{6}{|c|}{(SD)}				  &			-	         &	 4.1459   &		68.7468	\\	
												\hline
									\end{tabular}	
								\end{table}
								
				\vspace{-0.08cm}

\section{Conclusion}
In this paper, we introduced a novel code optimization technique which can be successfully applied to optimize degree distributions for both standard LDPC codes and MET-LDPC codes. We then proposed a joint optimization technique for MET-LDPC codes which allows the optimization of the MET structure and degree distribution given just the number of edge classes and maximum node degrees. 
We found that our proposed AR method works best for optimizing the edge degree distribution for a given set of allowed degrees while Dif.E works best for optimizing the set of allowed degrees.


\bibliographystyle{IEEEtran}

\end{document}